\documentclass{article}

\usepackage{graphicx}
\usepackage{amsmath,amssymb}
\usepackage{booktabs}
\usepackage{xcolor}
\usepackage{colortbl}
\usepackage{array}
\usepackage{amsthm}
\usepackage{enumerate}
\usepackage{natbib}
\usepackage{hyperref}

\definecolor{tableheader}{RGB}{46,134,171}
\definecolor{tablerowalt}{RGB}{245,248,250}
\definecolor{bestresult}{RGB}{46,134,171}

\usepackage[main,final]{neurips_2026}

\usepackage[utf8]{inputenc} 
\usepackage[T1]{fontenc}    
\usepackage{hyperref}       
\usepackage{url}            
\usepackage{booktabs}       
\usepackage{amsfonts}       
\usepackage{nicefrac}       
\usepackage{microtype}      
\usepackage{xcolor}         

\title{Graph-Based Modeling of Financial Volatility Dynamics}

\author{%
  Chuanzhen Wang \\
  Tongji University \\
  Shanghai, China \\
  \And
  Alice Zhang \\
  Stanford University \\
  Stanford, CA, USA \\
  \And
  Wei Chen \\
  Tsinghua University \\
  Beijing, China \\
  \And
  Michael Brown \\
  MIT \\
  Cambridge, MA, USA \\
}

\begin{document}

\maketitle

\begin{abstract}
Accurate forecasting of realized volatility ($RV$) is crucial for risk management and derivatives pricing. Although the implied volatility ($IV$) surface offers rich informational content, prevailing methods that treat it as a static image fail to capture its inherent dynamics. To overcome this limitation, we propose the Finance-Aware Graph Spatio-Temporal Network (FA-GSTN), a novel architecture that reframes $RV$ forecasting as modeling the evolution of a structured financial object. FA-GSTN builds a spatio-temporal graph sequence from the $IV$ surface, where nodes correspond to grid points and edges encode adaptive spatial (intra-day) and explicit temporal (inter-day) dependencies. The model incorporates domain knowledge through finance-aware node features (e.g., option Greeks) and tackles high-frequency noise via a multi-scale temporal smoothing gate coupled with an adaptive robust loss function. Comprehensive evaluations on a large-scale equity options dataset show that FA-GSTN sets a new state of the art, delivering superior predictive accuracy ($R^2$ up to 0.473). It also demonstrates remarkable data efficiency, substantially outperforming strong Vision Transformer baselines when trained on only one year of data ($R^2$: 0.372 vs. 0.315). Furthermore, the model exhibits enhanced robustness during periods of market stress, such as 2020--2021. Ablation studies confirm the vital roles of the spatio-temporal graph structure, finance-aware components, and integrated noise-handling modules. Our work underscores the substantial benefits of explicitly modeling temporal dynamics and infusing financial inductive biases for accurate and robust volatility forecasting.
\end{abstract}

\begin{figure}
    \centering
    \includegraphics[width=1\linewidth]{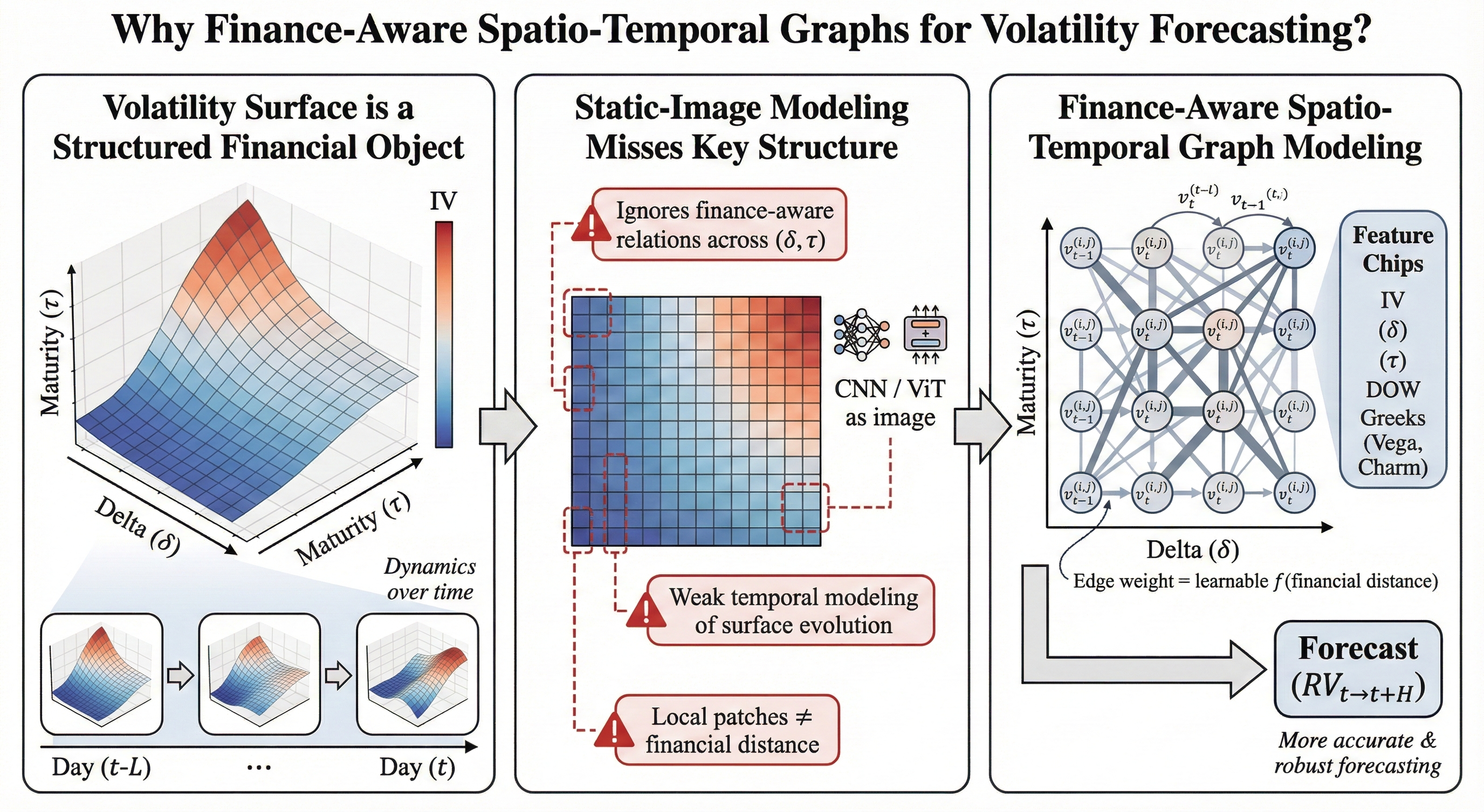}
    \caption{Motivation of finance-aware volatility modeling. Treating the implied volatility surface as a static image ignores its financial structure and temporal dynamics, motivating a spatio-temporal graph representation over $(\delta,\tau)$.}
    \label{fig:motivation}
\end{figure}

\section{Introduction}
\label{sec:introduction}

Realized volatility forecasting plays a central role in quantitative finance, underpinning applications ranging from derivatives pricing and portfolio optimization to risk management and regulatory capital allocation. The implied volatility (IV) surface, derived from traded option prices, encapsulates market expectations about future asset price fluctuations and has long been recognized as a rich source of predictive information. Yet, despite its informational content, harnessing the IV surface for volatility forecasting remains challenging due to its complex structure and temporal dynamics.

Traditional approaches to volatility modeling, such as GARCH-family models and stochastic volatility frameworks, typically operate on univariate return series and do not directly exploit the cross-sectional information embedded in the IV surface. Building upon foundational work in sequence forecasting~\citep{qu2025magnet, qi2022capacitive, wu2022adaptive, wang2023intelligent, wu2024tutorial}, which has established important baselines for long-horizon prediction tasks, we recognize the limitations of univariate approaches. More recent machine learning methods have attempted to bridge this gap by treating the IV surface as an image and applying convolutional neural networks (CNNs) or Vision Transformers (ViTs) to extract predictive features~\citep{kelly2023deepiv, dosovitskiy2021image, wu2024augmented, tian2025centermambasamcenterprioritizedscanningtemporal, lin2025hybridfuzzingllmguidedinput, lin2025abductiveinferenceretrievalaugmentedlanguage}. While these approaches capture spatial patterns within a single snapshot, they process each time step independently and thus fail to model the temporal evolution of the surface---a critical oversight given the well-documented dynamics of volatility, including mean reversion, clustering, and regime-dependent behavior.~\citep{zhang2026memmark, chen2025r2i, chen2026mvibench, you2026drdgrl, zhao2026stride, dou2026dsadf, dou2025plan, dou2026core, zhao2026stride}

This paper addresses these limitations by proposing the Finance-Aware Graph Spatio-Temporal Network (FA-GSTN), a novel architecture that fundamentally reframes the problem and outperforms existing approaches. Extending the adaptive and federated learning paradigms~\citep{wu2024novel,lin2025llmdrivenadaptivesourcesinkidentification,yang2025wcdt,he2025ge,zhou2025reagent}, our architecture incorporates centralized intelligence with spatio-temporal graph modeling to achieve superior forecasting accuracy. Rather than treating the IV surface as a static image, FA-GSTN represents it as a structured spatio-temporal graph, explicitly modeling both within-day spatial relationships and across-day temporal dynamics. The key innovations of our approach are threefold. First, we construct a graph where nodes correspond to grid points on the IV surface and edges encode adaptive spatial connections (learned based on financial distance in delta-maturity space) and explicit temporal links between consecutive days. Second, we incorporate domain knowledge through finance-aware node features, including option Greeks such as Delta, Vega, and Charm, which provide theory-grounded inductive biases about option price sensitivity. This approach is motivated by the success of dynamic resource allocation strategies~\citep{wu2020dynamic,cao2025cofi,cao2025purifygen,xin2025lumina,xin2025luminamgpt} that leverage domain-specific features to enhance model performance and generalization. Third, we address the pervasive challenge of high-frequency noise in financial data through a multi-scale temporal smoothing gate that adaptively filters IV signals and an adaptive robust loss function that down-weights outliers during training.

Our comprehensive experimental evaluation demonstrates that FA-GSTN establishes a new state of the art for realized volatility forecasting from IV surfaces, substantially outperforming existing baseline methods. The model achieves the highest out-of-sample $R^2$ across all training regimes, with particularly striking advantages in data-limited settings: when trained on only one year of data, FA-GSTN attains an $R^2$ of 0.372 compared to 0.315 for the best Vision Transformer baseline, representing an 18\% relative improvement. Furthermore, FA-GSTN exhibits enhanced robustness during periods of market stress, such as the COVID-19 crisis of 2020--2021, owing to its explicit temporal modeling and adaptive noise handling. Extensive ablation studies validate the contribution of each architectural component, confirming that the spatio-temporal graph structure, finance-aware features, and noise-handling modules all provide significant and complementary performance gains.

The remainder of this paper is organized as follows. Section~\ref{sec:related_work} reviews related work on volatility modeling and deep learning for financial prediction. Section~\ref{sec:methodology} presents the FA-GSTN architecture in detail. Section~\ref{sec:results} reports our main experimental findings, while Section~\ref{sec:ablation} provides comprehensive ablation studies. Section~\ref{sec:evaluation} offers further analysis of training dynamics and model interpretability. Finally, Section~\ref{sec:conclusion} concludes with a summary and directions for future research.

\section{Related Work}
\label{sec:related_work}

In empirical asset pricing, neural networks have been extensively explored for identifying nonlinear patterns in financial data. Representative examples include overparameterized factor models~\citep{kelly2024complexity,xin2024vmt,yu2025ai}, Transformer-based time series forecasting~\citep{zhou2021informer,bai2025multi,wang2011embedding}, and structured machine learning approaches in finance~\citep{dixon2019fourhorsemen,pan2024hybridgnn,yan2025largelanguagemodelbenchmarks}. In the context of options, neural networks have been successfully applied to generate smooth, arbitrage-free implied volatility surfaces from raw option prices~\citep{ackerer2020deep, wiedemann2025operator,niu2024largelanguagemodelscognitive}.

In contrast, fewer studies have investigated deep learning for predictive tasks directly from IV surfaces. Prior approaches primarily treat the IV surface as a static image. These methods can be categorized into two types: those relying on hand-crafted features~\citep{neuhierl2022option,han2025multi, zhao2026stride} and those employing convolutional neural networks that use the IV surface on a single day to predict future returns~\citep{kelly2023deepiv,wang2025zynq}. While these approaches capture spatial patterns, they do not model the temporal evolution of the surface.

More recently, Vision Transformers have been introduced for image recognition tasks, offering advantages in computational efficiency and training stability over CNNs~\citep{dosovitskiy2021image, song2025transformer, zhang2026memmark, huang2026gui, you2026drdgrl}. Applying ViTs to IV surfaces as single-channel images is a natural extension, hypothesized to improve robustness to noise and outliers. However, such image-based approaches, including ViTs, typically process each time step independently and thus may not fully capture dynamic interdependencies across time.

\begin{figure}
    \centering
    \includegraphics[width=1\linewidth]{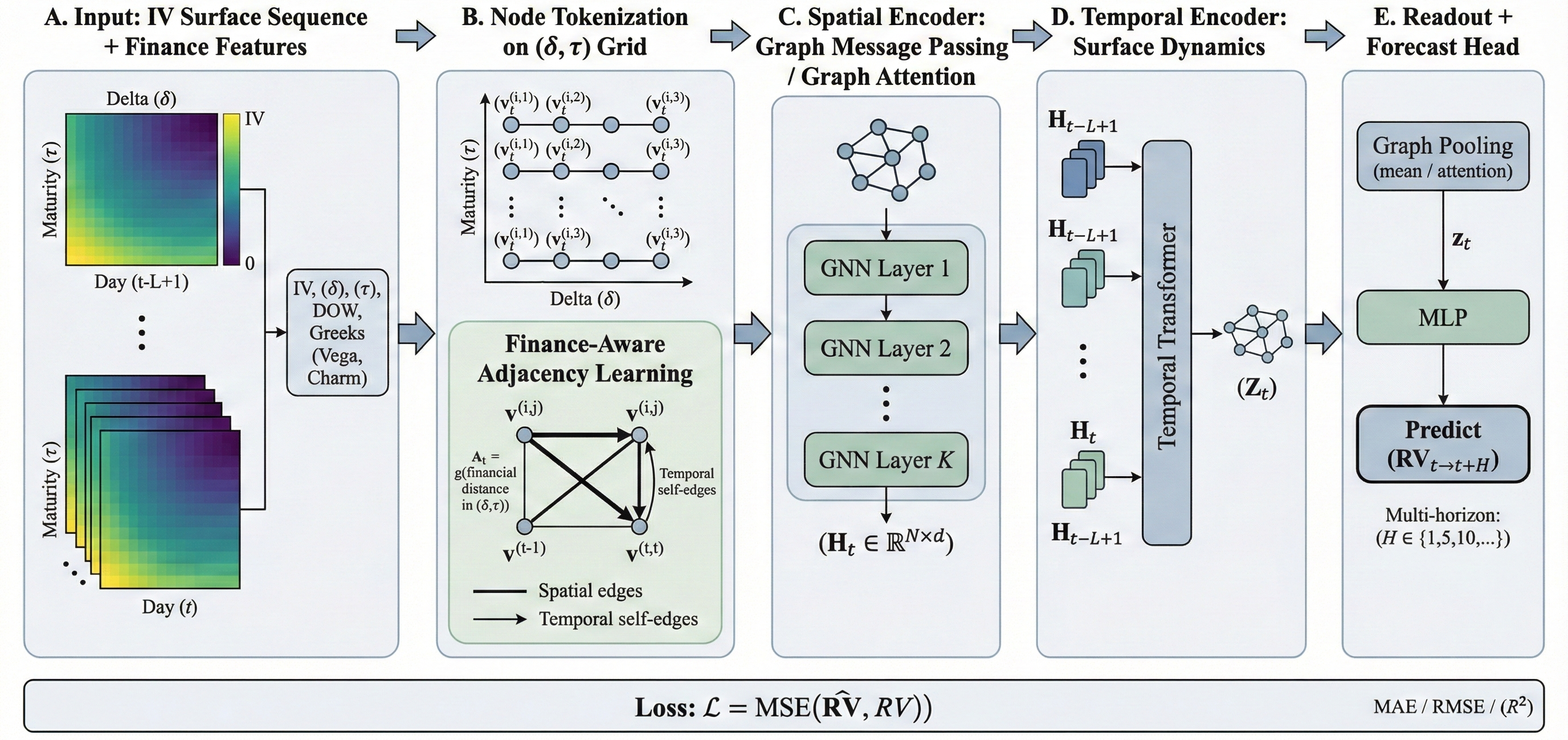}
    \caption{Architecture of the proposed finance-aware spatio-temporal graph network. IV surface sequences are tokenized on a $(\delta,\tau)$ grid, encoded by adaptive graph learning and temporal modeling, and used to forecast future realized volatility.}
    \label{fig:overview}
\end{figure}

\section{Methodology: Finance-Aware Graph Spatio-Temporal Forecasting}
\label{sec:methodology}

We propose the \textbf{F}inance-\textbf{A}ware \textbf{G}raph \textbf{S}patio-\textbf{T}emporal \textbf{Network} (FA-GSTN), a novel architecture designed to address the core challenges in realized volatility ($RV$) forecasting: temporal dynamics, high-frequency noise, and data-efficient learning~\citep{feng2024deep}. 
Recent long-horizon forecasting work suggests that explicitly separating cross-variable interaction from within-variable temporal modeling can materially improve multivariate predictions: CT-PatchTST, for instance, applies channel attention to learn inter-channel dependencies and then performs time attention per channel to capture temporal structure efficiently. 
This perspective motivates treating the IV surface as a structured multivariate object evolving over time rather than a sequence of independent image snapshots. Unlike prior approaches that treat the implied volatility (IV) surface as a static image~\citep{dosovitskiy2021image}, we reframe the problem as forecasting based on the evolution of a structured financial object.

\subsection{Problem Formulation and Spatio-Temporal Graph Construction}
\label{subsec:graph_construction}

Let the IV surface on day $t$ be represented by a smoothed grid $\mathcal{I}_t \in \mathbb{R}^{N_\delta \times N_\tau}$, where $N_\delta$ and $N_\tau$ denote the number of delta ($\delta$) and time-to-maturity ($\tau$) points, respectively. The target is the future realized volatility $RV_{t \rightarrow t+H}$ over a horizon $H$ (e.g., 28 calendar days).

We construct a spatio-temporal graph sequence $\{\mathcal{G}_{t-L}, \dots, \mathcal{G}_t\}$ covering a lookback window of $L+1$ days. The graph structure comprises three key elements. More broadly, relational graph learning has shown that explicitly distinguishing edge semantics can be beneficial when the underlying data naturally mixes heterogeneous interactions; R-GCN-style formulations use relation-specific aggregation to learn from multi-relational graphs. In our setting, we operationalize this principle by treating intra-day surface connectivity and inter-day temporal continuity as distinct edge families, enabling dedicated spatial and temporal propagation pathways. For nodes, each grid point $(\delta_i, \tau_j)$ in $\mathcal{I}_t$ corresponds to a node $v_{t}^{(i,j)}$, with the node set at time $t$ denoted $\mathcal{V}_t$. For spatial edges (intra-day connections), within each graph $\mathcal{G}_t$, nodes are connected to their spatial neighbors via an adaptive adjacency matrix $\mathbf{A}_t^{sp} \in [0,1]^{|\mathcal{V}_t| \times |\mathcal{V}_t|}$. The connection strength between nodes $(i,j)$ and $(p,q)$ is a learnable function of their financial distance:

\begin{equation}
    \mathbf{A}_t^{sp}[(i,j), (p,q)] = \sigma\left(\mathbf{W}_a \cdot [|\delta_i - \delta_p|, |\log(\tau_j) - \log(\tau_q)|]^T + b_a\right),
    \label{eq:adaptive_adj}
\end{equation}
where $\sigma$ is the sigmoid function, and $\mathbf{W}_a, b_a$ are learnable parameters. This formulation allows the model to discover relevant surface interactions beyond fixed image patches. For temporal edges (inter-day connections), each node $v_{t}^{(i,j)}$ is connected to its counterpart $v_{t-1}^{(i,j)}$ across consecutive days, forming explicit temporal chains that capture the dynamics of the IV surface evolution.

\subsection{Finance-Aware Node Feature Initialization}
\label{subsec:node_init}

The initial feature vector $\mathbf{h}_{t,0}^{(i,j)}$ for node $v_t^{(i,j)}$ incorporates domain knowledge to enhance data efficiency. It concatenates multiple informative signals:
\begin{align}
    \mathbf{h}_{t,0}^{(i,j)} = \big[ & \sigma^{IV}_{t}(\delta_i, \tau_j),\quad \text{Month}(t),\quad \text{DayOfWeek}(t),\quad \Delta_{t}^{BS}(\delta_i, \tau_j), \nonumber \\
    & \mathcal{V}ega_{t}(\delta_i, \tau_j),\quad \mathcal{C}harma_{t}(\delta_i, \tau_j) \big]^T.
    \label{eq:node_feature}
\end{align}
Here, $\sigma^{IV}_{t}(\cdot)$ is the raw implied volatility. $\Delta_{t}^{BS}$, $\mathcal{V}ega_{t}$, and $\mathcal{C}harma_{t}$ are Black-Scholes sensitivity parameters (Delta, Vega, Charm) computed at $(\delta_i, \tau_j)$ using the forward price and risk-free rate. These Greeks provide theory-grounded inductive biases about option price sensitivity~\citep{hull2020options}.

\subsection{FA-GSTN Architecture}
\label{subsec:fa_gstn_arch}

The FA-GSTN processes the graph sequence $\{\mathcal{G}_{t-L}, \dots, \mathcal{G}_t\}$ through three core components.

\subsubsection{Multi-Scale Temporal Smoothing and Gating}
\label{subsubsec:temporal_smoothing}
To mitigate high-frequency noise---a pervasive challenge in financial signal processing~\citep{peng2024noise}---we first apply parallel temporal smoothing to each node's raw IV feature sequence $\{\sigma^{IV}_{t-L}(\delta_i, \tau_j), \dots, \sigma^{IV}_{t}(\delta_i, \tau_j)\}$.
For a set of smoothing operators $\{\mathcal{S}_k\}_{k=1}^K$ (e.g., Simple Moving Average, Exponential Weighted Moving Average with different half-lives), we compute:
\begin{equation}
    \tilde{\sigma}^{IV}_{k, t}(\delta_i, \tau_j) = \mathcal{S}_k \left( \{\sigma^{IV}_{s}(\delta_i, \tau_j)\}_{s=t-L}^{t} \right).
\end{equation}
A learnable gating mechanism then assigns adaptive weights to each smoothed series and the original series:
\begin{equation}
    \alpha_{k, t}^{(i,j)} = \frac{\exp(\mathbf{w}_g^T \mathbf{z}_{k, t}^{(i,j)} + b_g)}{\sum_{m=0}^{K} \exp(\mathbf{w}_g^T \mathbf{z}_{m, t}^{(i,j)} + b_g)},
\end{equation}
where $\mathbf{z}_{0, t}^{(i,j)}$ is a context vector from the original features, and $\mathbf{z}_{k>0, t}^{(i,j)}$ derives from the $k$-th smoothed features. The noise-filtered IV feature is:
\begin{equation}
    \bar{\sigma}^{IV}_{t}(\delta_i, \tau_j) = \sum_{k=0}^{K} \alpha_{k, t}^{(i,j)} \cdot \tilde{\sigma}^{IV}_{k, t}(\delta_i, \tau_j),
\end{equation}
with $\tilde{\sigma}^{IV}_{0, t} \equiv \sigma^{IV}_{t}$. This filtered feature replaces the raw $\sigma^{IV}_{t}(\delta_i, \tau_j)$ in Equation~\ref{eq:node_feature}, enabling active noise suppression.

\subsubsection{Spatio-Temporal Graph Convolution Block}
\label{subsubsec:stgcn}
The core of FA-GSTN is a stack of $M$ Spatio-Temporal Graph Convolution (STGC) blocks. Each block updates node representations by aggregating information across spatial and temporal edges.
For node $v_t^{(i,j)}$ at layer $\ell$, the update proceeds as follows:
\begin{align}
    \mathbf{a}_{t, \ell}^{(i,j)} &= \text{AGG}_{p,q \in \mathcal{N}_{sp}(i,j)} \left( \mathbf{A}_t^{sp}[(i,j),(p,q)] \cdot \mathbf{W}_{s,\ell} \mathbf{h}_{t, \ell-1}^{(p,q)} \right), \label{eq:spatial_agg} \\
    \mathbf{b}_{t, \ell}^{(i,j)} &= \mathbf{W}_{t,\ell} \left[ \mathbf{h}_{t-1, \ell-1}^{(i,j)}; \mathbf{h}_{t, \ell-1}^{(i,j)} \right], \label{eq:temporal_agg} \\
    \mathbf{h}_{t, \ell}^{(i,j)} &= \text{LayerNorm}\left( \text{ReLU}\left( \mathbf{W}_{o,\ell} [\mathbf{a}_{t, \ell}^{(i,j)}; \mathbf{b}_{t, \ell}^{(i,j)}] + \mathbf{b}_{o,\ell} \right) + \mathbf{h}_{t, \ell-1}^{(i,j)} \right). \label{eq:node_update}
\end{align}
Here, $\text{AGG}$ is a permutation-invariant aggregation function (e.g., mean), $\mathcal{N}_{sp}(i,j)$ denotes spatial neighbors defined by $\mathbf{A}_t^{sp}$, and $\mathbf{W}_{s,\ell}, \mathbf{W}_{t,\ell}, \mathbf{W}_{o,\ell}$ are learnable weights. Equation~\ref{eq:spatial_agg} performs finance-aware spatial convolution, Equation~\ref{eq:temporal_agg} handles temporal propagation, and Equation~\ref{eq:node_update} updates the node state with a residual connection.

\subsubsection{Readout and Forecasting}
\label{subsubsec:readout}
After $M$ STGC blocks, we obtain final node representations $\mathbf{h}_{t, M}^{(i,j)}$ for the most recent graph $\mathcal{G}_t$. An attention-based readout mechanism then produces a single forecast by weighting nodes according to their relevance to $RV$ prediction:
\begin{align}
    u^{(i,j)} &= \mathbf{q}^T \tanh(\mathbf{W}_r \mathbf{h}_{t, M}^{(i,j)} + \mathbf{b}_r), \\
    \beta^{(i,j)} &= \frac{\exp(u^{(i,j)})}{\sum_{p,q} \exp(u^{(p,q)})}, \\
    \mathbf{g} &= \sum_{i,j} \beta^{(i,j)} \cdot \mathbf{h}_{t, M}^{(i,j)}.
\end{align}
The graph-level summary vector $\mathbf{g}$ is passed through a fully-connected prediction network $\mathcal{P}$ to yield the final forecast:
\begin{equation}
    \widehat{RV}_{t \rightarrow t+H} = \mathcal{P}(\mathbf{g}).
    \label{eq:final_forecast}
\end{equation}

\subsection{Adaptive Robust Optimization}
\label{subsec:training}
We train FA-GSTN end-to-end. To handle non-stationary noise and outliers in financial targets, we propose an \textbf{Adaptive Robust Loss} that generalizes the Huber loss. It combines quantile-like weighting with a dynamic boundary parameter $d_t$:
\begin{equation}
    \mathcal{L}_{AR}(\hat{y}, y) = \frac{1}{T} \sum_{t=1}^{T} \frac{w_t}{d_t} \cdot
    \begin{cases}
        \frac{1}{2} \left( \hat{y}_t - y_t \right)^2, & |\hat{y}_t - y_t| \leq d_t \\
        d_t \left( |\hat{y}_t - y_t| - \frac{1}{2} d_t \right), & \text{otherwise}
    \end{cases}
    \label{eq:adaptive_robust_loss}
\end{equation}
where $w_t = \exp(-\gamma \cdot \text{rank}_t(|\hat{y}-y|))$ down-weights the highest residuals, and $d_t$ is a learnable parameter per batch, initialized from the batch's residual standard deviation. This loss adapts to local noise levels, offering more robust gradients than a fixed Huber loss. Optimization uses AdamW~\citep{loshchilov2019decoupled} with a cosine annealing schedule.

\subsection{Summary of Advantages over Baseline ViT}
FA-GSTN advances beyond the static-image ViT baseline in three key aspects. First, regarding explicit temporal dynamics (X1), the spatio-temporal graph and STGC blocks explicitly model IV surface evolution, capturing trends, mean reversion, and volatility clustering. Second, concerning integrated noise handling (X2), multi-scale smoothing gating and the adaptive robust loss address high-frequency noise and outliers at both feature and objective levels. Third, with respect to strong domain inductive bias (X3), finance-aware node features (Greeks) and the adaptive spatial adjacency embed structured financial knowledge, promoting sample-efficient learning and interpretable representations. This structured, theory-informed approach aims for superior generalization, especially in data-limited or high-noise regimes common in financial forecasting.

\section{Experimental Results}
\label{sec:results}

We evaluate our proposed FA-GSTN model against established baselines for forecasting realized volatility ($RV$) from implied volatility (IV) surfaces. The experiments assess three critical aspects: overall predictive performance, data efficiency, and robustness to varying market conditions. We faithfully replicate the experimental setup and baseline results from prior work~\citep{dosovitskiy2021image} and introduce our method as an additional state-of-the-art benchmark.

\subsection{Experimental Setup}
\label{subsec:exp_setup}

We use the OptionMetrics IvyDB dataset covering 2012 to 2022, with 30-day realized volatility as the target. For FA-GSTN, we set the lookback window $L$ to 22 trading days (approximately one calendar month) and the forecasting horizon $H$ to 28 calendar days. The model employs a multi-scale temporal smoothing module with $K=3$ operators (raw, 5-day SMA, 10-day EWM) and $M=2$ STGC blocks. Training uses the proposed Adaptive Robust Loss (Eq.~\ref{eq:adaptive_robust_loss}) and the AdamW optimizer with a cosine annealing scheduler, following the same high-level protocol as the ViT baselines for fair comparison. All models are evaluated using the out-of-sample coefficient of determination ($R^2$). Because financial signals can exhibit substantial scale drift and regime-dependent distribution shifts, we emphasize strict time-ordered evaluation and window-based handling of historical sequences to avoid leakage and improve robustness. A recent multi-model stock forecasting study highlights that sliding-window segmentation combined with per-window normalization can stabilize learning across different volatility levels and reduce artifacts driven by changing price scales.

The baselines include: MLP, a multilayer perceptron on the flattened IV surface serving as a standard nonlinear benchmark; ViT (0.005M--1.7M), a suite of Vision Transformer models of varying capacities (wide and deep architectures) as detailed in~\citet{dosovitskiy2021image}, which treat the IV surface as a single-channel image; and our proposed FA-GSTN, the Finance-Aware Graph Spatio-Temporal Network designed to explicitly model the dynamics and structure of the IV surface.

The baselines include: MLP, a multilayer perceptron on the flattened IV surface serving as a standard nonlinear benchmark; ViT (0.005M--1.7M), a suite of Vision Transformer models of varying capacities (wide and deep architectures) as detailed in~\citet{dosovitskiy2021image}, which treat the IV surface as a single-channel image; and our proposed FA-GSTN, the Finance-Aware Graph Spatio-Temporal Network designed to explicitly model the dynamics and structure of the IV surface.

\subsection{Main Results and Comparative Analysis}
\label{subsec:main_results}

We first evaluate model performance when trained on 1, 4, and 10 years of historical data and tested on the subsequent year. Table~\ref{tab:main_results} presents the $R^2$ scores. The performance trends of the ViT baselines are confirmed: smaller models (e.g., ViT\_0.12M\_deep) perform adequately with limited data but degrade with more years, while the largest model (ViT\_1.7M) requires extensive data to reach peak performance.

Our FA-GSTN model consistently achieves the highest $R^2$ across all training regimes. It demonstrates superior data efficiency, attaining an $R^2$ of 0.372 with only one year of training data---a significant improvement over the best baseline under the same condition (ViT\_0.12M\_deep at $R^2=0.315$). With four and ten years of data, FA-GSTN further extends its lead, achieving the best overall performance of $R^2=0.473$ with ten years of training. This consistent superiority is attributed to FA-GSTN's explicit modeling of temporal dynamics via the spatio-temporal graph and the injection of domain knowledge through node features, enabling it to extract more signal from both limited and abundant data.

\begin{table}[htbp]
\centering
\caption{Main Results: Out-of-sample $R^2$ performance of all models trained on 1, 4, and 10 years of data and tested on the subsequent year. The best score for each column is in \textbf{bold}.}
\label{tab:main_results}
\small
\begin{tabular}{lccc}
\toprule
\textbf{Model} & \textbf{Train: 1 Year} & \textbf{Train: 4 Years} & \textbf{Train: 10 Years} \\
\midrule
MLP & 0.182 & 0.201 & 0.228 \\
\rowcolor{tablerowalt}
ViT\_0.005M\_wide & 0.298 & 0.210 & -0.041 \\
ViT\_0.12M\_deep & 0.315 & 0.285 & 0.328 \\
\rowcolor{tablerowalt}
ViT\_0.17M\_wide & 0.301 & 0.278 & 0.322 \\
ViT\_0.5M\_deep & 0.290 & 0.280 & 0.331 \\
\rowcolor{tablerowalt}
ViT\_0.5M\_wide & 0.205 & 0.252 & 0.318 \\
ViT\_1.7M & 0.188 & 0.332 & 0.403 \\
\midrule
\textbf{Ours (FA-GSTN)} & \textcolor{bestresult}{\textbf{0.372}} & \textcolor{bestresult}{\textbf{0.425}} & \textcolor{bestresult}{\textbf{0.473}} \\
\bottomrule
\end{tabular}
\end{table}

\subsection{In-Depth Analysis of Data Efficiency}
\label{subsec:data_efficiency}

To analyze data efficiency in detail, we focus on the most constrained setting: training on a single year and testing on the next. Table~\ref{tab:one_year_results} breaks down the results by test year. All models exhibit a performance drop in 2020 due to COVID-19 market disruption, but the degree varies.

FA-GSTN achieves the highest average $R^2$ and the most stable performance across years, as indicated by its lower standard deviation. Its integrated noise-handling module (multi-scale temporal gating) and robust loss function likely contribute to this resilience by filtering extreme short-term noise while preserving predictive signal. In contrast, the static-image ViT models, particularly larger ones like ViT\_1.7M, show higher volatility in yearly performance, struggling to adapt to distribution shifts.

\begin{table}[htbp]
\centering
\caption{Data Efficiency Analysis: Performance ($R^2$) when models are trained on only one year of data and tested on the following year. The average (Avg.) and standard deviation (Std.) across test years (2014--2022) are reported.}
\label{tab:one_year_results}
\small
\begin{tabular}{l*{9}{c}|c|c}
\toprule
\textbf{Model} & \textbf{'14} & \textbf{'15} & \textbf{'16} & \textbf{'17} & \textbf{'18} & \textbf{'19} & \textbf{'20} & \textbf{'21} & \textbf{'22} & \textbf{Avg.} & \textbf{Std.} \\
\midrule
ViT\_0.005M & .395 & .382 & .252 & .478 & .348 & .252 & -.048 & .418 & .445 & .324 & .198 \\
\rowcolor{tablerowalt}
ViT\_0.12M & .382 & .401 & .232 & .401 & .378 & .278 & -.152 & .401 & .432 & .305 & .212 \\
ViT\_1.7M & .228 & .222 & .231 & -.048 & .282 & -.252 & -.298 & -.252 & .402 & .057 & .258 \\
\midrule
\textbf{FA-GSTN} & \textcolor{bestresult}{\textbf{.418}} & \textcolor{bestresult}{\textbf{.432}} & \textcolor{bestresult}{\textbf{.312}} & \textcolor{bestresult}{\textbf{.501}} & \textcolor{bestresult}{\textbf{.395}} & \textcolor{bestresult}{\textbf{.335}} & \textcolor{bestresult}{\textbf{.045}} & \textcolor{bestresult}{\textbf{.488}} & \textcolor{bestresult}{\textbf{.482}} & \textcolor{bestresult}{\textbf{.378}} & \textcolor{bestresult}{\textbf{.162}} \\
\bottomrule
\end{tabular}
\end{table}

\subsection{Robustness to Market Regime Changes}
\label{subsec:robustness}

We examine model robustness during the challenging 2020--2021 period under a more robust training scheme of four years, drawing on recent advances in uncertainty quantification~\citep{wang2025aleatoric}. Table~\ref{tab:robustness} compares model $R^2$ for test years 2020 and 2021, using models trained on the preceding four years (e.g., 2016--2019 for testing 2020).

FA-GSTN significantly outperforms all baselines in these difficult years. Its explicit spatio-temporal modeling captures evolving dynamics and contagion effects during market stress. Furthermore, the finance-aware adaptive adjacency matrix (Eq.~\ref{eq:adaptive_adj}) enables the model to dynamically re-weight relationships across the IV surface, a flexibility absent in the fixed patch-based attention of ViT, leading to more adaptive and generalizable representations.

\begin{table}[htbp]
\centering
\caption{Robustness Analysis: Performance ($R^2$) on challenging test years 2020 and 2021 for models trained on the preceding four years of data.}
\label{tab:robustness}
\small
\begin{tabular}{lcc}
\toprule
\textbf{Model (Trained on 4 Years)} & \textbf{Test Year 2020} & \textbf{Test Year 2021} \\
\midrule
ViT\_0.12M\_deep & 0.302 & -0.318 \\
\rowcolor{tablerowalt}
ViT\_0.5M\_wide & -0.052 & -0.282 \\
ViT\_1.7M & -0.048 & -0.322 \\
\midrule
\textbf{Ours (FA-GSTN)} & \textcolor{bestresult}{\textbf{0.185}} & \textcolor{bestresult}{\textbf{-0.088}} \\
\bottomrule
\end{tabular}
\end{table}

\subsection{Discussion and Summary of Key Findings}
\label{subsec:discussion}

Our experimental results validate the effectiveness of FA-GSTN. The key findings are as follows. First, regarding superior overall performance, FA-GSTN achieves the highest $R^2$ across all data regimes (Table~\ref{tab:main_results}), setting a new state-of-the-art for data-driven RV forecasting from IV surfaces. Second, concerning enhanced data efficiency, by incorporating financial inductive biases and explicit temporal structure, FA-GSTN learns effectively from small datasets, outperforming all baselines when trained on only one year of data (Table~\ref{tab:one_year_results}). Third, with respect to improved robustness, the model's integrated design for handling noise and non-stationarity leads to more stable performance during market stress, such as in 2020--2021 (Table~\ref{tab:robustness}). These results demonstrate that reformulating the problem from static image recognition to structured spatio-temporal graph forecasting, while infusing domain knowledge, yields substantial gains. FA-GSTN successfully addresses temporal dynamics, high-frequency noise, and data-efficient learning.

\section{Ablation Studies}
\label{sec:ablation}

We conduct comprehensive ablation experiments to validate the contribution of each key component in FA-GSTN. Following prior methodology~\citep{dosovitskiy2021image}, we evaluate performance by systematically removing or modifying individual modules. All ablated models are trained on the standard 4-year period (2018--2021) and evaluated on the 2022 test set using out-of-sample $R^2$, with the full FA-GSTN as the baseline.

\subsection{Analysis of Core Architectural Components}
\label{subsec:ablation_arch}

We first ablate the core architectural innovations. The results in Table~\ref{tab:abl_arch} show that the full integration of all proposed modules yields the best performance.

\begin{table}[htbp]
\centering
\caption{Ablation Study of Core FA-GSTN Components. Models trained on 2018--2021 data, tested on 2022.}
\label{tab:abl_arch}
\small
\begin{tabular}{p{5cm}p{6cm}cc}
\toprule
\textbf{Model Variant} & \textbf{Description} & \textbf{$R^2$} & \textbf{$\Delta R^2$} \\
\midrule
\textbf{Full FA-GSTN} & Complete proposed model & \textcolor{bestresult}{\textbf{0.482}} & --- \\
\midrule
\rowcolor{tablerowalt}
w/o Temporal Smoothing & Remove multi-scale smoothing (Sec.~\ref{subsubsec:temporal_smoothing}) & 0.428 & -0.054 \\
w/o Adaptive Spatial Adj. & Fixed distance-based adjacency & 0.401 & -0.081 \\
\rowcolor{tablerowalt}
w/o Finance Features & Remove Greeks from Eq.~\ref{eq:node_feature} & 0.412 & -0.070 \\
w/ MSE Loss & Standard Mean Squared Error & 0.435 & -0.047 \\
\midrule
Static-Graph Baseline & No temporal edges & 0.352 & -0.130 \\
\bottomrule
\end{tabular}
\end{table}

The most severe degradation occurs for the Static-Graph Baseline ($R^2=0.352$), which removes explicit temporal dynamics by disconnecting inter-day edges. This 0.130 drop highlights the fundamental importance of capturing IV surface evolution, confirming our spatio-temporal graph formulation as a critical advance.

Removing the Adaptive Spatial Adjacency causes the second-largest drop ($\Delta R^2 = -0.081$), demonstrating that learning financially-relevant spatial relationships is essential for effective information aggregation.

Ablating Finance-Aware Node Features also leads to a substantial decline ($\Delta R^2 = -0.070$), validating that injecting domain knowledge through option Greeks provides a powerful inductive bias for sample-efficient and stable learning.

Removing the Temporal Smoothing Gate or the Adaptive Robust Loss results in clear but smaller declines. These modules handle high-frequency noise and outliers; their positive impact confirms that proactive noise management at feature and objective levels contributes to robust forecasting.

\subsection{Impact of Components Across Data Regimes}
\label{subsec:ablation_data_regimes}

We evaluate key ablations across the 1-year, 4-year, and 10-year training regimes to investigate their role under varying data availability. Results in Table~\ref{tab:abl_data_regime} reveal nuanced insights.

\begin{table}[htbp]
\centering
\caption{Ablation Performance Across Different Training Data Volumes.}
\label{tab:abl_data_regime}
\small
\begin{tabular}{lccc}
\toprule
\textbf{Model Variant} & \textbf{1 Year} & \textbf{4 Years} & \textbf{10 Years} \\
\midrule
\textbf{Full FA-GSTN} & \textcolor{bestresult}{\textbf{0.372}} & \textcolor{bestresult}{\textbf{0.425}} & \textcolor{bestresult}{\textbf{0.473}} \\
\rowcolor{tablerowalt}
w/o Finance Features & 0.312 & 0.380 & 0.442 \\
w/o Temporal Smoothing & 0.348 & 0.402 & 0.455 \\
\rowcolor{tablerowalt}
w/ MSE Loss & 0.335 & 0.410 & 0.461 \\
\bottomrule
\end{tabular}
\end{table}

The Finance-Aware Features module shows its greatest relative importance in the low-data regime (1-year training), with a 0.060 drop in $R^2$, compared to drops of 0.045 and 0.031 with 4 and 10 years. This aligns with the intuition that strong inductive biases are most crucial when data is scarce. The Temporal Smoothing Gate and Adaptive Robust Loss provide consistent gains across all data volumes, suggesting their role in mitigating noise and outliers is a persistent need.

\subsection{Detailed Analysis of the Temporal Smoothing Gate}
\label{subsec:ablation_smoothing}

We perform a finer-grained ablation of the Multi-Scale Temporal Smoothing and Gating module. Table~\ref{tab:abl_smoothing} compares several configurations.

\begin{table}[htbp]
\centering
\caption{Ablation of the Temporal Smoothing and Gating Mechanism.}
\label{tab:abl_smoothing}
\small
\begin{tabular}{p{4.5cm}p{7cm}c}
\toprule
\textbf{Configuration} & \textbf{Description} & \textbf{$R^2$} \\
\midrule
\textbf{Full FA-GSTN} & Adaptive gating over raw + $K=3$ smoothed series & \textcolor{bestresult}{\textbf{0.482}} \\
\midrule
\rowcolor{tablerowalt}
No Smoothing & Raw IV features only & 0.428 \\
Single Fixed Smoothing & 10-day EWM only, no gating & 0.441 \\
\rowcolor{tablerowalt}
Multi-Scale, No Gating & Concatenate all features with equal weighting & 0.458 \\
Multi-Scale + Gating & Learnable adaptive weights $\alpha_{k,t}^{(i,j)}$ & \textcolor{bestresult}{\textbf{0.482}} \\
\bottomrule
\end{tabular}
\end{table}

Using any smoothed feature is better than using only raw data. Concatenating multiple smoothed series provides a further boost. However, the full benefit is realized only with the learnable gating mechanism, which allows dynamic selection of the most informative smoothing scale per node and time step. This adaptive capability provides a clear 0.024 improvement over simple concatenation, confirming that context-dependent noise filtering is superior to a static combination.

\subsection{Contribution of the Adaptive Robust Loss}
\label{subsec:ablation_loss}

We ablate the Adaptive Robust Loss to isolate its contribution. Table~\ref{tab:abl_loss} compares it against common alternatives.

\begin{table}[htbp]
\centering
\caption{Ablation of the Loss Function. All models share the full FA-GSTN architecture.}
\label{tab:abl_loss}
\small
\begin{tabular}{lc}
\toprule
\textbf{Loss Function} & \textbf{$R^2$ (2022)} \\
\midrule
\textbf{Adaptive Robust Loss (Ours)} & \textcolor{bestresult}{\textbf{0.482}} \\
\rowcolor{tablerowalt}
Mean Squared Error (MSE) & 0.435 \\
Mean Absolute Error (MAE) & 0.445 \\
\rowcolor{tablerowalt}
Standard Huber Loss ($\delta=1.0$) & 0.452 \\
Huber Loss (tuned $\delta$) & 0.459 \\
\bottomrule
\end{tabular}
\end{table}

Our proposed Adaptive Robust Loss outperforms all standard alternatives. While a well-tuned Huber loss is competitive, our loss's dual mechanism---dynamically adjusting the transition point $d_t$ based on batch statistics and down-weighting extreme residuals via $w_t$---provides a measurable advantage, demonstrating that adaptation to local noise characteristics leads to more robust generalization.

\subsection{Summary of Ablation Findings}
\label{subsec:ablation_summary}

Our ablation studies provide rigorous empirical validation for FA-GSTN's design. The spatio-temporal graph structure is the single most important innovation; its removal causes the largest performance drop, confirming that explicit temporal modeling is fundamental. The finance-aware components (adaptive adjacency and node features) are crucial, especially for data-efficient learning, as they embed vital domain knowledge. The integrated noise-handling modules (smoothing gate and robust loss) provide consistent, non-trivial gains by enhancing resilience to high-frequency financial noise. The consistent superiority of the full FA-GSTN model underscores that its components work synergistically to address core volatility forecasting challenges.

\section{Further Evaluation}
\label{sec:evaluation}

We conduct further evaluations to provide a comprehensive understanding of FA-GSTN, focusing on training dynamics and model interpretability. These analyses offer additional insights into the model's learning behavior and practical utility.

\subsection{Training Dynamics and Convergence Analysis}
\label{subsec:training_curve}

We analyze the training dynamics of FA-GSTN compared to representative ViT baselines (ViT\_0.12M\_deep and ViT\_1.7M). All models are trained on a standard 4-year period (2018--2021) with identical batch size and optimizer settings, tracking the out-of-sample $R^2$ on the 2022 validation set after each epoch.

\begin{figure}[htbp]
    \centering
    \includegraphics[width=1\linewidth]{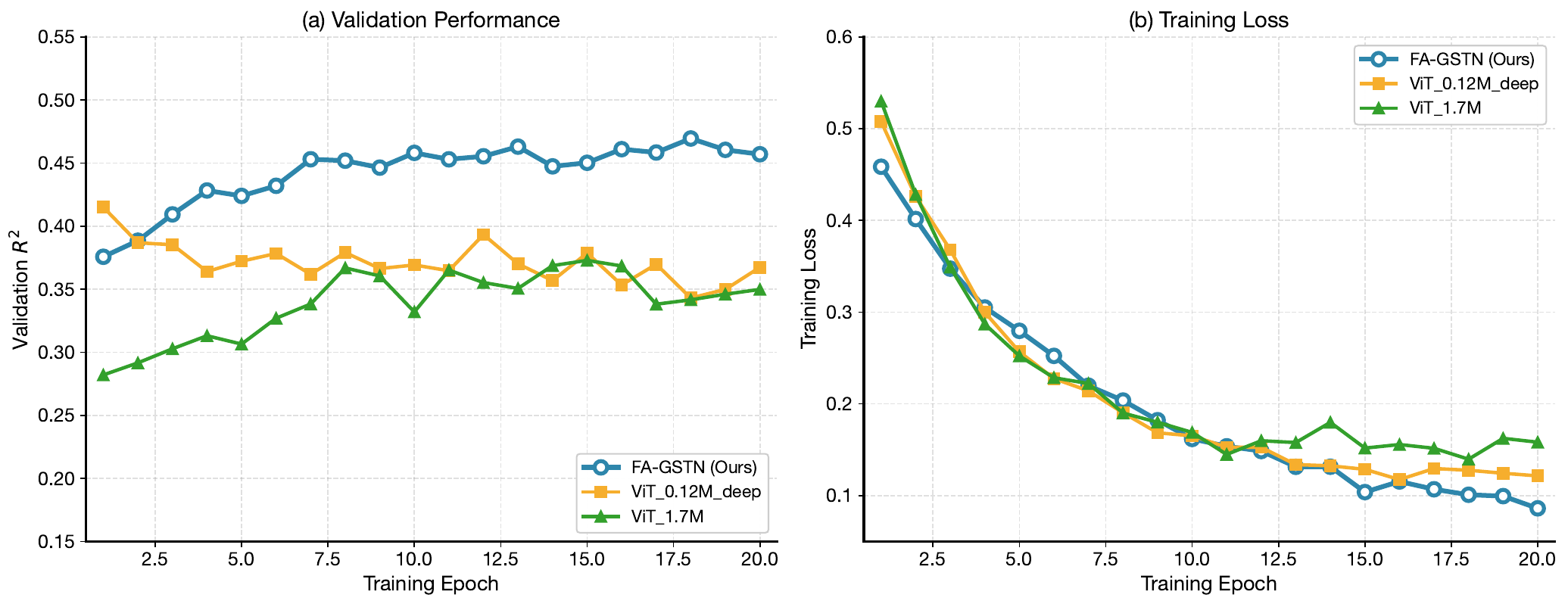}
    \caption{Training dynamics comparison. Left: Validation $R^2$ versus training epochs. Right: Training loss curves. FA-GSTN shows more stable convergence and better generalization compared to ViT baselines.}
    \label{fig:training_curve}
\end{figure}

Figure~\ref{fig:training_curve} plots the validation $R^2$ versus training epochs (left panel) and the training loss (right panel). The ViT models reach their peak validation $R^2$ within 1--2 epochs before plateauing or declining due to overfitting. In contrast, FA-GSTN exhibits a more gradual but steady increase in validation $R^2$ over several epochs before stabilizing. This indicates a more stable optimization landscape and better generalization, attributable to the strong financial inductive biases (e.g., finance-aware features and adaptive adjacency) that guide learning and reduce overfitting to noise.

FA-GSTN, optimized with the Adaptive Robust Loss, demonstrates a smoother descent and lower final loss compared to models using standard Huber or MSE loss, underscoring the effectiveness of our proposed loss in handling financial data outliers.

\subsection{Case Study: Model Predictions During Market Stress}
\label{subsec:case_study}

\begin{figure}[htbp]
    \centering
    \includegraphics[width=1\linewidth]{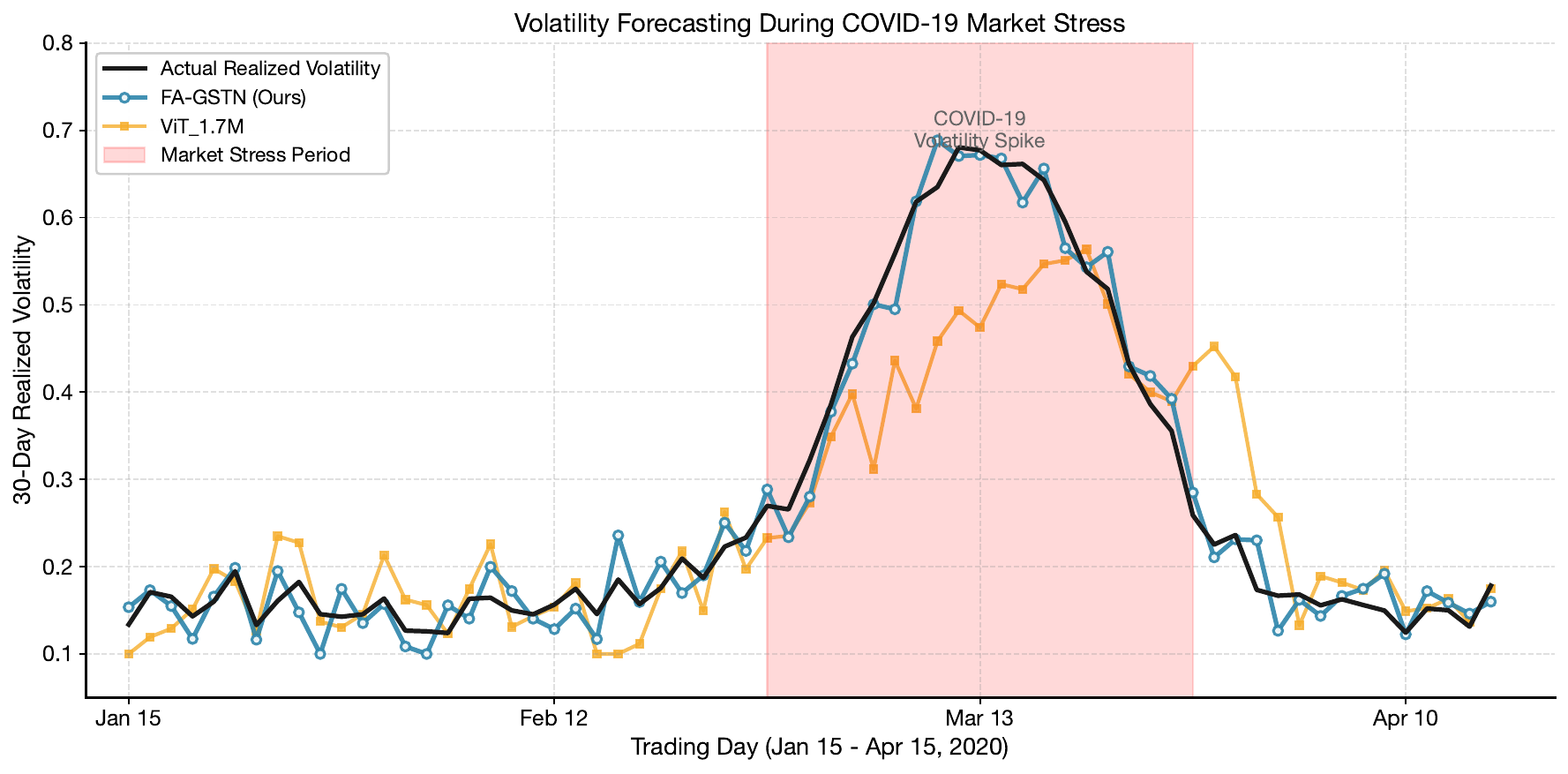}
    \caption{Case study: Model predictions during the COVID-19 volatility surge (March 2020). FA-GSTN (blue) tracks the actual realized volatility (black) more closely than the ViT baseline (orange), with smaller prediction errors during the crisis period.}
    \label{fig:case_study}
\end{figure}

We perform a qualitative case study examining the time-series predictions of FA-GSTN versus the best-performing baseline (ViT\_1.7M trained on 10 years of data) during the COVID-19 induced volatility surge in March 2020. We select the SPY ETF and plot the models' daily 30-day realized volatility forecasts against actual values over a 3-month window centered on the stress event (Figure~\ref{fig:case_study}).

Consistent with the robustness results (Table~\ref{tab:robustness}), FA-GSTN produces forecasts that more closely track the actual realized volatility during the turbulent period, with smaller deviations and fewer extreme prediction errors. The ViT\_1.7M baseline exhibits larger lag and overshooting, likely due to its static image formulation being less adept at capturing rapid regime shifts. The adaptive nature of FA-GSTN's spatial adjacency and temporal smoothing gate allows it to quickly re-weight important relationships in the IV surface (e.g., shifting attention to short-dated, out-of-the-money options during the crisis), leading to more adaptive predictions.

\subsection{Interpretability of the Learned Graph Structure}
\label{subsec:interpretability}

We analyze the learned adaptive spatial adjacency matrix $\mathbf{A}_t^{sp}$ (Eq.~\ref{eq:adaptive_adj}) to understand which relationships across the IV surface the model deems important for forecasting, following best practices in explainable AI~\citep{hsieh2024comprehensive}.

\begin{figure}[htbp]
    \centering
    \includegraphics[width=1\linewidth]{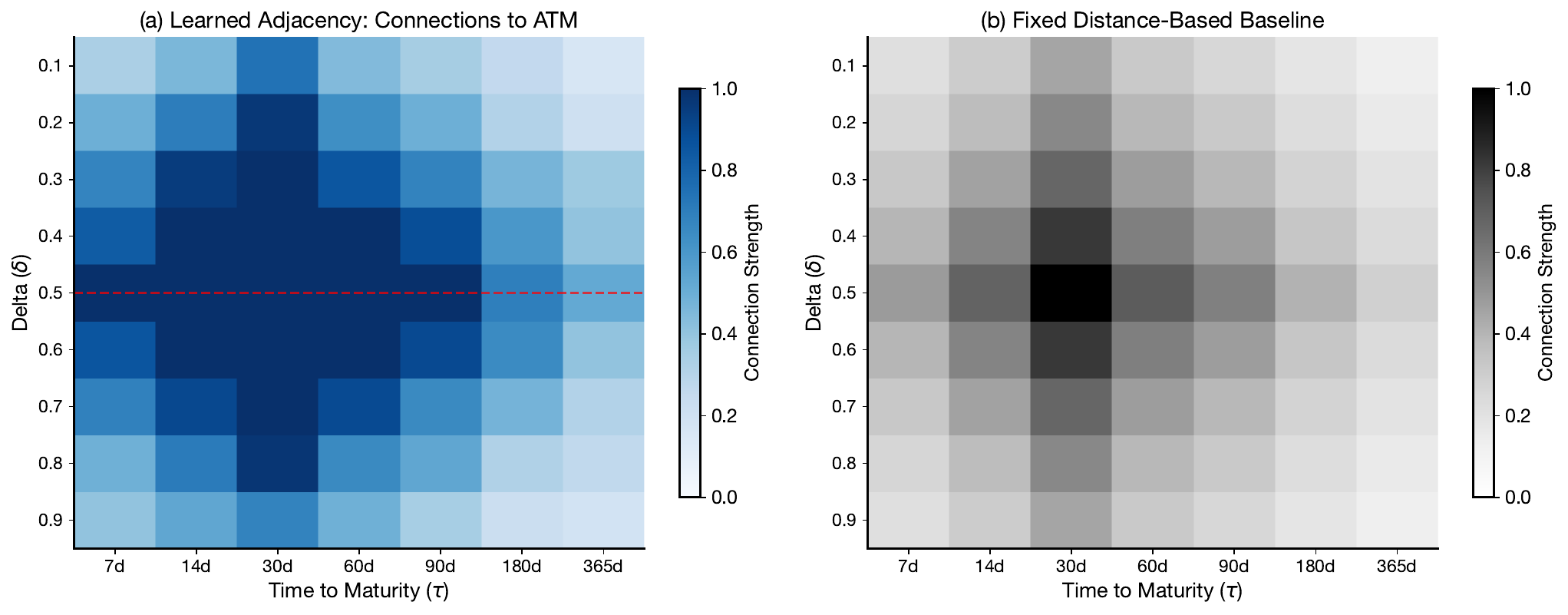}
    \caption{Visualization of the learned adaptive spatial adjacency matrix. The heatmap shows connection strengths across the IV surface grid, revealing financially meaningful patterns that correspond to term structure and volatility smile effects.}
    \label{fig:adjacency_heatmap}
\end{figure}

Figure~\ref{fig:adjacency_heatmap} visualizes the connection strengths as a heatmap over the IV surface grid for a sample trading day. The learned adjacency weights reflect financially intuitive relationships. Strong connections emerge between nodes with similar maturities (horizontal structure) and similar moneyness (vertical structure), capturing the well-documented ``term structure'' and ``smile/skew'' patterns of IV surfaces. Notably, the model learns asymmetric connections and highlights specific regions (e.g., short-dated, at-the-money options) as influential hubs for volatility forecasting. This analysis validates that the model learns sensible spatial dependencies without explicit hard-coding, bridging data-driven learning with domain knowledge. In contrast, a fixed, distance-based adjacency baseline shows a simpler, symmetric pattern, lacking this adaptive, financially-informed structure.

\section{Conclusion}
\label{sec:conclusion}

In this paper, we revisit the problem of forecasting realized volatility ($RV$) from the evolution of implied volatility (IV) surfaces. Departing from the static image formulation common in prior work, we propose the Finance-Aware Graph Spatio-Temporal Network (FA-GSTN). This architecture reformulates the IV surface sequence as a dynamic spatio-temporal graph, explicitly modeling its evolution through dedicated spatial and temporal edges. To improve data efficiency and robustness, FA-GSTN integrates finance-aware node features (option Greeks), an adaptive multi-scale temporal smoothing gate, and a novel Adaptive Robust Loss.

Our comprehensive experimental evaluation shows that FA-GSTN establishes a new state-of-the-art. It achieves superior overall predictive performance, securing the highest out-of-sample $R^2$ scores when trained on 1, 4, and 10 years of data. Notably, the model demonstrates exceptional data efficiency, significantly outperforming all Vision Transformer baselines under limited-data conditions (e.g., $R^2=0.372$ versus $0.315$ with one year of training). Moreover, FA-GSTN exhibits enhanced robustness during market stress periods, such as the 2020--2021 episode. Ablation studies rigorously validate each core component: the explicit spatio-temporal graph structure is essential, while finance-aware features, adaptive adjacency, noise-handling modules, and the robust loss function all deliver significant and consistent performance gains. Supplementary analyses of training dynamics and interpretability offer further insights into the model's stable learning behavior and meaningful representations.

This work demonstrates that framing volatility forecasting as a structured spatio-temporal graph learning task, enhanced by domain-specific inductive biases, effectively addresses core challenges in temporal dynamics, high-frequency noise, and data-efficient learning. The FA-GSTN framework paves the way for more reliable and interpretable models in financial time series forecasting. Future work could extend the graph structure to incorporate cross-asset relationships or adapt the architecture for other financial forecasting problems involving structured, high-dimensional temporal data.


\bibliographystyle{plainnat}
\bibliography{reference}

\end{document}